\documentclass[12pt,preprint]{aastex}

\begin{document}

%\slugcomment{}

\title{Detection of a Third Planet in the HD 74156 System Using the Hobby-Eberly Telescope\footnotemark}

\author{Jacob L. Bean\altaffilmark{2,3}, Barbara E. McArthur\altaffilmark{2}, G. Fritz Benedict\altaffilmark{2}, \& Amber Armstrong\altaffilmark{2}}

\footnotetext[1]{Based on data obtained with the Hobby-Eberly Telescope (HET). The HET is a joint project of the University of Texas at Austin, the Pennsylvania State University, Stanford University, Ludwig-Maximilians-Universit\"{a}t Muenchen, and Georg-August-Universit\"{a}t G\"{o}ttingen. The HET is named in honor of its principal benefactors, William P. Hobby and Robert E. Eberly.}

\altaffiltext{2}{Dept.\ of Astronomy and McDonald Observatory, University of Texas, 1 University Station, C1402, Austin, TX 78712}

\altaffiltext{3}{Now at the Institut f\"{u}r Astrophysik, Georg-August-Universit\"{a}t G\"{o}ttingen, Friedrich-Hund-Platz 1, 37077 G\"{o}ttingen, Germany; bean@astro.physik.uni-goettingen.de}

\begin{abstract}
We report the discovery of a third planetary mass companion to the G0 star HD 74156. High precision radial velocity measurements made with the Hobby-Eberly Telescope aided the detection of this object. The best fit triple Keplerian model to all the available velocity data yields an orbital period of 347 days and minimum mass of 0.4 $M_{Jup}$ for the new planet. We determine revised orbital periods of 51.7 and 2477 days, and minimum masses of 1.9 and 8.0 $M_{Jup}$ respectively for the previously known planets. Preliminary calculations indicate that the derived orbits are stable, although all three planets have significant orbital eccentricities ($e$ = 0.64, 0.43, and 0.25). With our detection, HD 74156 becomes the eighth normal star known to host three or more planets. Further study of this system's dynamical characteristics will likely give important insight to planet formation and evolutionary processes.

\end{abstract}


\section{INTRODUCTION}
More than 200 planets have been discovered around nearby stars other than the Sun using the radial velocity method\footnote{A regularly updated list of reported extrasolar planets is maintained at the the Extrasolar Planets Encyclopaedia website \url{http://exoplanet.eu}.}. Despite the high number of known exoplanets, the hunt for additional planets around nearby stars with the radial velocity method remains a fundamental research area. Planet candidates around nearby stars are the most suitable for followup study like photometric monitoring for transits \citep[e.g.][]{henry00, charbonneau00, castellano00, bouchy05, sato05, shankland06, lopez06, gillon07}, astrometric perturbation measurements \citep[e.g.][]{benedict02, mcarthur04, benedict06, bean07}, thermal emission searches \citep[e.g.][]{deming05, harrington06, richardson07, knutson07, cowan07}, attempted direct imaging detection \citep[e.g.][]{janson07}, and dynamical characterization \citep[e.g.][]{laughlin99, chiang02, adams06, ford05, barnes06}. The results of these types of investigations improve our understanding of planet formation and evolution. 

Radial velocity planet searches are also a crucial component of the quest to determine the uniqueness of our own solar system. It should be noted that only recently have some surveys achieved the necessary precision and timespan to detect true Jupiter analogs (i.e. planets with roughly the same mass and orbital semimajor axis), and it is not yet possible to detect analogs of any of the other planets in our solar system. Therefore, the continuation and improvement of radial velocity planet searches should be a top priority.

After our successful detection of an additional companion to $\rho^{1}$ Cnc \citep[= 55 Cnc,][]{mcarthur04}, we began a small radial velocity monitoring program in 2004 with goal of finding additional planets in already known planetary systems. Our methodology was to take advantage of the Hobby-Eberly Telescope's (HET) queue scheduled operations to obtain high cadence radial velocities of selected systems. In addition, we chose to obtain multiple radial velocity measurements per observational epoch to reduce the impact of both statistical and stellar noise on the search for previously hidden planets. This intensive technique limits the number of possible targets, but is a natural complement to the larger planet search programs. 

In this paper we report the first result from this project, the detection of a third planet around HD 74156. In \S 2 we give the properties and background information on the star. In \S 3 we describe our radial velocity measurements with the HET. We present our analysis that yields the detection of the new planet in \S 4. We conclude with a discussion of our result in \S 5.

\section{HD 74156 PROPERTIES}
According to the \textit{Hipparcos} catolog \citep{esa97}, HD 74156 (HIP 42723) is a G0 star about 65 pc from the Sun ($\pi_{abs} = 15.49 \pm 1.10$ mas) with visual magnitude $V$ = 7.61 and color $\bv$ = 0.59. \citet{naef04} previously reported the discovery of two planetary mass companions to the star based on its radial velocity variations. The planets had estimated minimum masses $M_{b} \sin i = 1.9\ M_{Jup}$ and $M_{c} \sin i = 6.2\ M_{Jup}$ assuming the primary's mass $M_{A} = 1.27\ M_{\sun}$. The lower mass planet is in a short period orbit ($P_{b}$ = 51.6 day), while the higher mass planet has a longer orbital period ($P_{c}$ = 2025 day). Both planets' orbits were reported as highly eccentric ($e$ = 0.64 and 0.58). 

\citet{vf05} determined the iron abundance [Fe/H] = +0.13 for HD 74156. This result agrees with that from \citet{santos03}, who found [Fe/H] = +0.15. Therefore, it is metal rich compared to the solar neighborhood, but average among host stars to high mass planets \citep{fv05}. 

\citet{naef04} suggested that HD 74156 might be slightly evolved based on its large estimated luminosity relative to its spectral type. \citet{takeda07} report an age of 3.7 $\pm$ 0.4 Gyr based on a comparison of the star's properties with theoretical models. This age estimate implies that HD 74156 is still on the main sequence, although ages for field stars are notoriously difficult to determine. \citet{naef04} reported seeing no emission in the Ca \textsc{ii} H line core, which is consistent with the star having at least an intermediate age. 

\citet{takeda07} also proposed a mass $M_{A} = 1.24 \pm 0.04\ M_{\sun}$ for HD 74156. This is consistent with the mass adopted in the discovery paper ($M_{A} = 1.27\ M_{\sun}$), which was estimated by \citet{santos03}. We elected to adopt the former value in this paper for consistency with our previous work. We included the uncertainty suggested by \citet{takeda07} in the calculation of the planet minimum masses given in \S 4.

\section{RADIAL VELOCITY MEASUREMENTS}
We made high precision radial velocity measurements of HD 74156 using the iodine absorption cell method \citep[e.g.][]{butler96}. The specific details of our implementation are the same as for our previous work described in \citet{bean07}. 

Observations using the HET to feed the High Resolution Spectrograph \citep[HRS,][]{tull98} were carried out by on-site telescope operators during 82 nights between UT dates 2004 December 3 and 2007 May 11. The HRS was used in the resolution R = 60,000 mode with a 316 line mm$^{-1}$ echelle grating. The cross dispersion grating was positioned so that the central wavelength of the order that fell in the break between the two CCD chips was 5936 \AA. A temperature controlled cell containing molecular iodine gas (I$_{2}$) was inserted in front of the spectrograph slit entrance during all exposures to imprint lines that provided a contemporaneous wavelength scale and instrumental profile fiducial for the high precision radial velocity measurements. The exposure times were nominally 150 s, but varied up to three times that occasionally to account for increased seeing and/or cloud cover. Three sequential exposures were obtained each night and a total of 242 spectra of HD 74156 for radial velocity measurements were collected. 

HD 74156 was also observed once on 2006 December 14 without the iodine cell and with the same instrument setup, but in the $R$ = 120,000 mode. The exposure time was 650 s. The spectrum from this observation served as a template for the radial velocity measurements.

CCD reduction and optimal spectral extraction were carried out for all the individual spectra using the REDUCE package \citep{piskunov02} and nightly calibration data. Relative radial velocities were measured from the spectra using the modeling algorithm that is described in detail by \citet{bean07}. We used the \textit{Gaussfit} program \citep{jefferys88} to reduce each of the 82 multiple observation sets to a single velocity. The velocities we measured are relative to an arbitrary zero point, and we determined an offset simultaneously with our orbit analysis to adjust them to the system barycenter (see \S 4). The final velocities with this offset subtracted are given in Table~\ref{tab:table1}. They have a median uncertainty of 2.6 m s$^{-1}$.

\section{ORBIT ANALYSIS}
We combined our measured radial velocities for HD 74156 with those from \citet[][``ELODIE'' and ``CORALIE'' samples]{naef04} to create a dataset that spans 9.33 years. No other high precision radial velocities for HD 74156 have been published. The timespan, number of data points, and RMS residuals to the final best fit model for each of the three individual velocity samples are given in Table~\ref{tab:table2}. The ELODIE and CORALIE measurements overlap, but there is a gap of 1.19 years between the final CORALIE measurement and our first measurement with the HET. 

We fit the total radial velocity dataset with a double Keplerian model to determine the two previously known planets' period, $P$, velocity semiamplitude, $K$, eccentricity, $e$, longitude of periastron, $\omega$, and time of periastron, $T_{P}$. To account for the heterogeneous nature of the dataset, we also simultaneously determined offset parameters for each individual velocity sample to adjust them to be relative to the system barycenter. We performed the fitting using the \textit{Gaussfit} program \citep{jefferys88} with both robust and least squares (``reduced chi squared'' metric, $\chi_{\nu}^{2}$) estimation to determine the parameter values that gave best match between our model and the measured data. For our robust fitting we used the ``fair'' metric described by \citet{rey83} with an adopted asymptotic relative efficiency (ARE) of 0.92. We refer to the models found with the two methods as ``Robust'' and ``Least Squares'' hereafter. In the case of the two planet model, the parameters determined using the two methods were consistent.

\begin{figure}
\epsscale{0.53}
\plotone{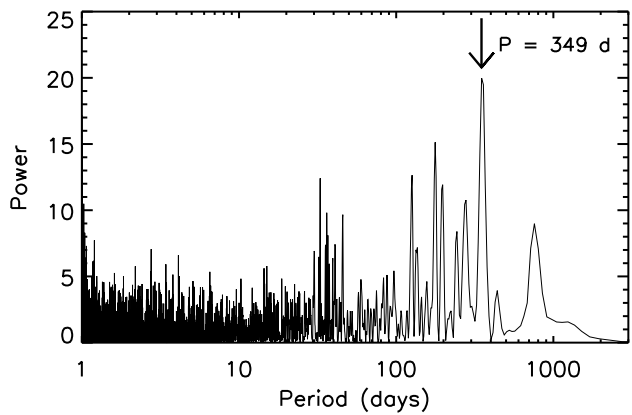}
\caption{Periodogram of the velocity residuals from the two planet Robust model. The spike around 349 days indicates a remaining periodic signal that is not accounted for by the previously known planets. The false alarm probability of the peak is 0.0014\%.}
\label{fig:f1}
\end{figure}

To search for evidence of additional planets in the HD 74156 system we calculated a periodogram \citep{press92} of the velocity residuals from the two planet model fit. The result is shown in Figure~\ref{fig:f1}. The highest peak in the periodogram is at a period around 349 days and has a formal false alarm probability (FAP) of 0.0014\%. We used the bootstrap method described by \citet{endl02} to further investigate the likelihood that this signal was real. We calculated a periodogram and noted the maximum power for each of 10,000 simulated data sets. The data sets were generated by randomly scrambling the two planet model residuals while maintaining the sampling of the observations. We did not find a maximum power in any of the trial periodograms as high or higher than the peak at 349 days in the periodogram of the real two planet model residuals (19.98). This implies a FAP for the 349 day signal of less than 0.01\%, which is consistent with the formal estimate. We took this as evidence for a real periodic signal in the radial velocities not accounted for by the two previously known planets. 

We explored the possibility that the detected periodic signal is attributable to a third planet in the system by fitting a triple Keplerian model to the radial velocity dataset. We again used both the robust and least squares estimation methods. The quality of the fits improved significantly with the addition of the third component. However, we found that the RMS of the HET velocity residuals (6.0 m s$^{-1}$) remained more than twice the median of the measurement uncertainties (2.6 m s$^{-1}$). This discrepancy could be due to using an incomplete model of the system (i.e. there are more than 3 planets), intrinsic variations of the star's photosphere that mimic real radial velocity changes, improper weighting of the ELODIE and CORALIE data, and/or errors in our velocity measurement method. Using the method of \citet{wright05}, \citet{butler06} estimated the radial velocity ``jitter'' for HD 74156 to be 4.0 m s$^{-1}$. Adding this value in quadrature with the HET velocity uncertainties yields errors that are slightly above what would be expected from the fit residuals and the fit $\chi_{\nu}^{2}$ = 0.88. This is not surprising because our method of taking multiple measurements over 10 -- 25 minutes and combining them into one measurement reduces the noise arising from short-term stellar variability. We ultimately decided to increase the HET uncertainties by a factor of two in order to account for potential errors in the data and to avoid overweighting them in the orbital fitting, but we cannot discount the possibility that there are additional planets in the system causing the higher than expected dispersion. Therefore, we have elected to publish our original estimate of the HET velocity uncertainties in Table~\ref{tab:table1} to avoid biasing future investigations.

We repeated both the two and three planet orbit fitting using the revised uncertainties for the HET data. The orbital parameters and associated uncertainties that we determined for the three planet models and minimum masses calculated from these parameters are given in Table~\ref{tab:table3}. We found that the $\chi_{\nu}^{2}$ of the Least Squares model improved from 2.3 to 1.4 by introducing the third planet into the model. The RMS of the HET velocity residuals from the fits improved from 8.5 to 6.0 m s$^{-1}$ for both methods. The RMS for the ELODIE residuals went from 12.7 to 10.8 m s$^{-1}$ in the Least Squares model and from 12.8 to 9.5 m s$^{-1}$ in the Robust model. However the RMS of the CORALIE residuals degraded from 10.3 to 10.5 m s$^{-1}$ in the Least Squares model and from 10.2 to 11.1 m s$^{-1}$ in the Robust model. The three planet model does yield an improved fit to the combined ELODIE and CORALIE dataset, with the RMS of the residuals from the Robust model improving from 11.7 to 10.3 m s$^{-1}$.

\begin{figure}
\epsscale{1.1}
\plotone{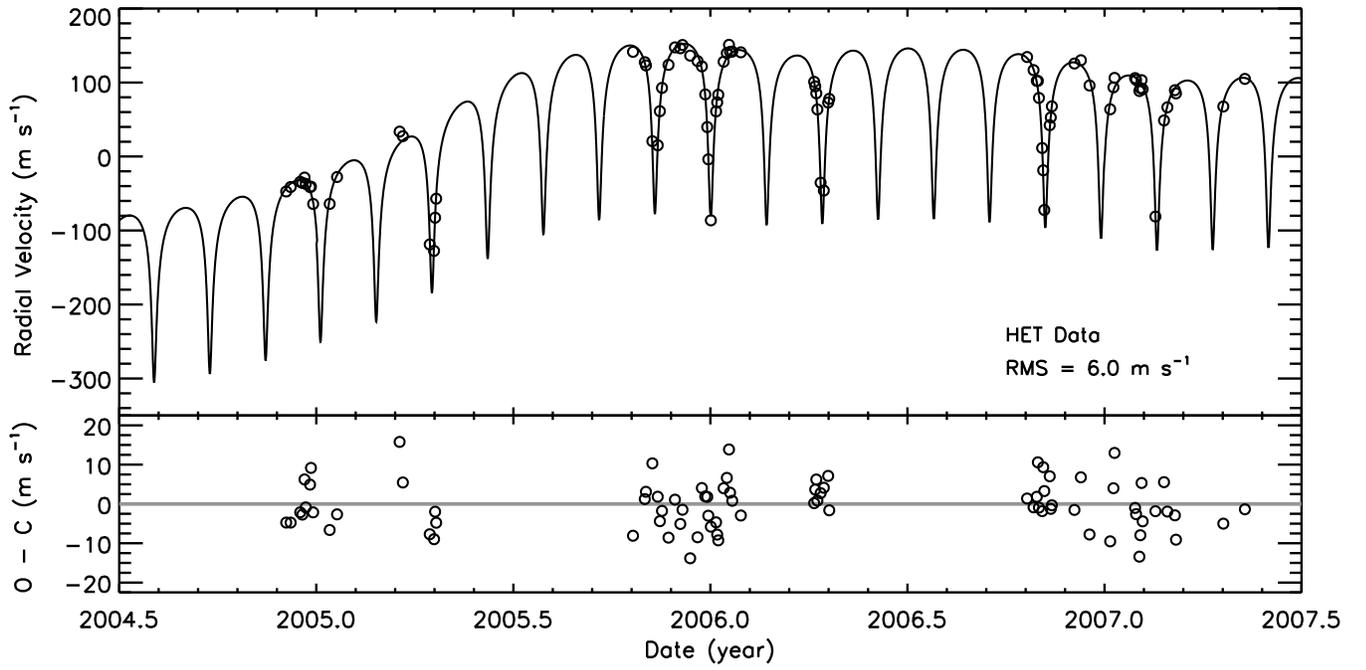}
\caption{Top: HET radial velocities (circles) and the Robust three planet model to all the data (line). Bottom: Residuals from the fit (circles). The error bars are omitted for clarity.} 
\label{fig:f2}
\end{figure}

\begin{figure}
\epsscale{0.513}
\plotone{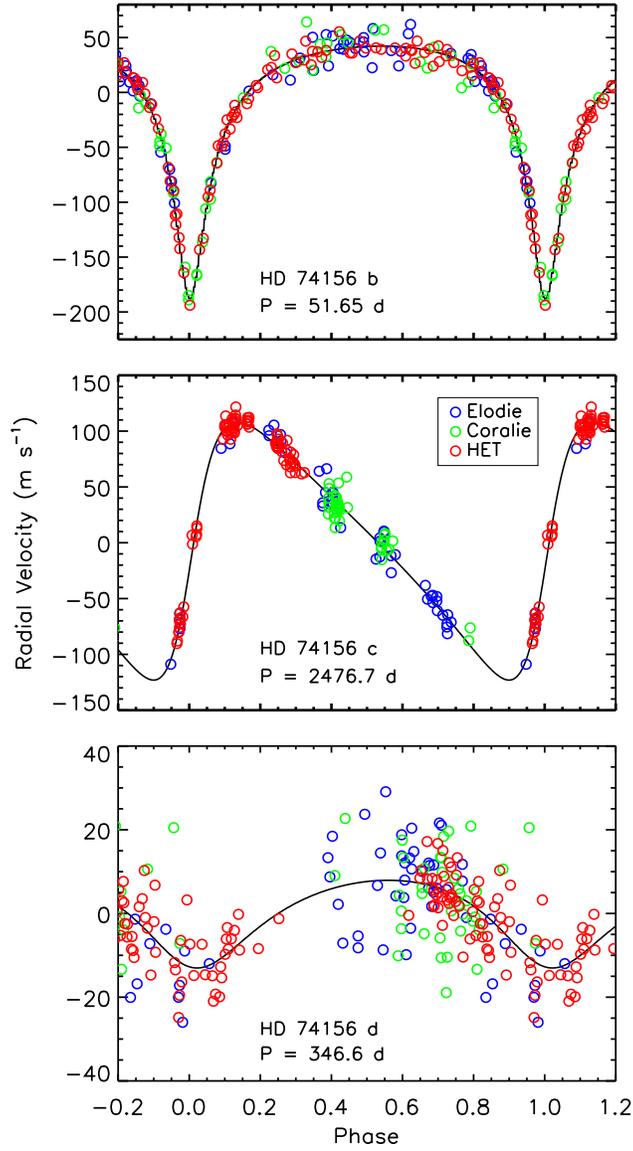}
\caption{Representation of the Robust three planet model. The radial velocity data (circles) are phased to each component's period with the other two components' orbits subtracted and the component's model velocities (lines). The error bars are omitted for clarity.}
\label{fig:f3}
\end{figure}

\begin{figure}
\epsscale{0.53}
\plotone{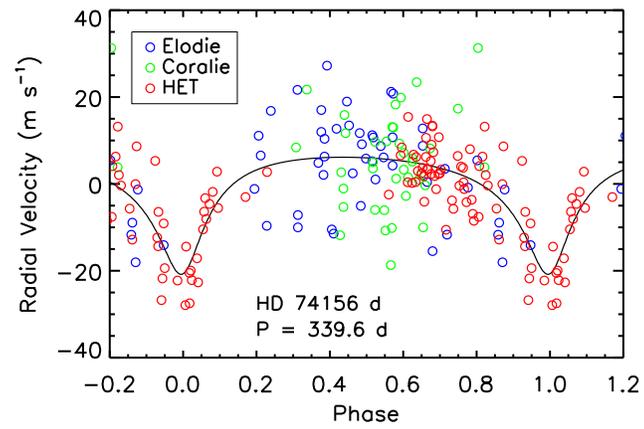}
\caption{Representation of the Least Squares orbit for the third planet (line) with the other two components' orbits subtracted from the data (circles). Note that the small differences in the parameters for the two other planets do yield slightly different residuals when their orbits are subtracted from the data. The error bars are omitted for clarity.}
\label{fig:f4}
\end{figure}

The HET velocities and the Robust fit are shown in Figure~\ref{fig:f2}. The radial velocity data is shown phased to each component's period and with the other two components' orbits subtracted assuming the Robust model in Figure~\ref{fig:f3}. The Robust and Least Squares models for the two previously known planets are very similar. However, the period, time of periastron, and eccentricity for the candidate third planet determined using the two different methods are inconsistent. The most notable difference between the models for this planet is in the derived eccentricities, with the Robust model having $e$ = 0.25 and the Least Squares model having $e$ = 0.55. For comparison, the new planet's phased orbit for the Least Squares model is shown in Figure~\ref{fig:f4}. 

The least squares method of fitting involves weighting the data points according to their input variances and is based on the assumption that the data are normally distributed. The purpose of using robust estimation instead is to reduce the influence of non-normal outliers in the data. Rather than attempting to identify outliers by hand, robust metrics place outlier identification and their subsequent reweighting on a rigorous footing. Data points with fit residuals larger than a certain threshold are iteratively down-weighted, rather than completely ignored, during the fitting process. However, the identification of some data points as outliers does not mean they are necessarily bad. It only means that they don't agree with adopted physical and noise models and thus should not carry their normal weight when assessing the fit quality to avoid avoid biasing the final solution. For example, a small signal due to a fourth planet in the system could cause some data points to seem highly discrepant. In this case the data points would be correct, but would adversely affect the fitting of a model that only includes three componenets if they were not identified and down-weighted. 

The similarity of the derived orbital parameters for the two previously known planets when using the robust and least squares methods is due to the fact that even the worst possible outliers in the radial velocity data are small relative to the amplitudes of the planets' velocity signals. However, the size of the outliers is the same or even larger than the amplitude of the velocity signal from the possible third planet. Therefore, whether potential outliers are identified and treated differently significantly influences the orbital solution for this component. 

The Robust model for HD 74156 d is based on the estimation that some of the HET data points are actually contaminating outliers. This isn't surprising given the variety of physical processes that can influence a measured stellar radial velocity at the 2 -- 3 m s$^{-1}$ level. The down-weighting of the identified outliers leads to a less eccentric orbit as the fit to highly deviant points with small input uncertainties is relaxed. The Least Squares model is the best fit to the data assuming that the input errors are scaled correctly and the data are normally distributed. The weight given to all the HET data points is based on their input uncertainties, which are twice those given in Table~\ref{tab:table1}. The result is a more eccentric orbit for the new planet because the solution is strongly influenced by the potential outliers. We favor the Robust model over the Least Squares model, but we chose to present and discuss both in order to illustrate the uncertainty in derived orbital parameters for planets when the signal is near the detection threshold in the data.

\begin{figure}
\epsscale{0.53}
\plotone{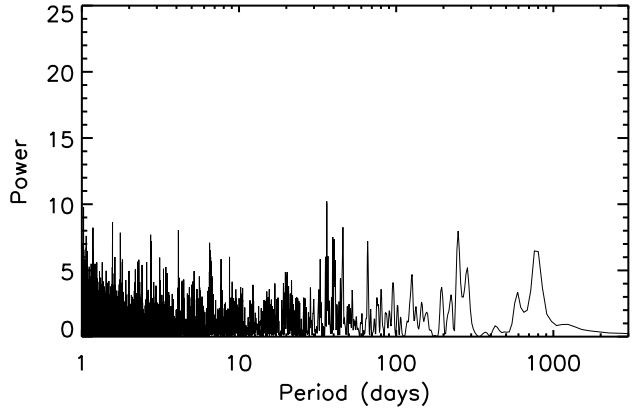}
\caption{Periodogram of the residuals to the three planet Robust model. No periodicity is detected with false alarm probability less than 22\%.}
\label{fig:f5}
\end{figure}

A periodogram of the three planet Robust model residuals is shown in Figure~\ref{fig:f5}. The periodogram of the residuals to the Least Squares model is essentially the same. The addition of the third component to the model removes the periodicity detected at 349 days. No periodicity is detected in the three planet model residuals with a FAP less than 22\%. This indicates that a summation of three Keplerian orbits adequately accounts for the detected periodic signals in the radial velocities,

We note that the periodogram of the two planet model residuals in Figure~\ref{fig:f1} also exhibits moderately significant power around 175 days. This is roughly half the value of the period we explored and ultimately determined for the potential third planet. Also, the 175 day spike in the two planet model residual periodogram has a smaller power (FAP = 0.09\% from the bootstrap simulation) than the spike at 349 days and is not present in the periodogram of the three planet model residuals. We therefore suspect that the smaller, moderately significant spikes around 175 days in the periodogram of the two planet model residuals were aliases. Nevertheless, we attempted to fit a three planet model to the radial velocity data where the third component had a period around 175 days. We were not able to determine a physically realistic set of parameters for the third component with this approach and various trials with fixed parameters yielded much worse fits than the three planet fits where the third component had a period close to 349 days ($\Delta\chi^{2} >$ 80 for the least squares approach).

The significant improvement in fit quality for both the Robust and Least Squares models and comparison of the two planet and three planet model residual periodograms lead us to propose that there is a third planet in the HD 74156 system with orbital parameters similar to those given in Table~\ref{tab:table3}. Beyond the detection of this new planet, our data and analysis has also yielded improved orbital parameters for HD 74156 c. We find for this planet a period 22\% longer and eccentricity reduced by 0.13 from the values determined by \citet{naef04}. This is not an unusual result when determining revised orbital parameters for long period planets with the addition of significantly more radial velocity data.

From our determined Robust model parameters we estimate $M_{b} \sin i = 1.88 \pm 0.03$, $M_{c} \sin i = 8.03 \pm 0.12$, and $M_{d} \sin i = 0.40 \pm 0.02\ M_{Jup}$ assuming $M_{A} = 1.24 \pm 0.04\ M_{\sun}$ \citep{takeda07}. The derived velocity semiamplitude for the new planet ($\sim$ 11.5 m s$^{-1}$) is roughly the same magnitude as the ELODIE and CORALIE residuals, while it is roughly twice that of the HET residuals. Therefore, the HET velocities are the most sensitive to the third planet. Its period is also close to a year (but significantly different enough so that it cannot be due an alias or error in the data) and the 2.5 years of HET observations do not give full phase coverage. More high precision observations over the next few years will be needed to refine the orbital parameters for this planet.

The eccentricities we have determined by fitting the radial velocity data indicate that the two previously known planets are in highly eccentric orbits, while the new planet is in at least a moderately eccentric orbit. This raises the question of whether the derived orbital parameters represent a stable three planet system. We made a preliminary check of this using the Runge Kutta numerical integrator in the \textit{Systemic Console}\footnote{Available at \url{http://www.oklo.org}. We used the May 30, 2007 version of the code.}. We integrated the planet positions forward in time for 1000 years assuming their true masses were equal to the minimum masses and that their orbits were in the same plane. We used a 0.05 day time step for the calculations. The 1000 year integration represents about 7100 full orbits of the innermost planet. We found that the semi-major axes of all three planets were not predicted to change from their initial values during the entirety of both integrations. Therefore, both sets of orbital parameters that we have determined do not describe highly unstable systems.

Prior to our discovery, \citet{raymond05} had determined that an additional Saturn-mass planet in the HD 74156 system having an orbital semimajor axis $a$ = 0.9--1.4 AU, and $e <$ 0.15 would very likely be dynamically stable for 100 million years. They had even predicted that because such a planet could exist, then it must exist based on their ``Packed Planetary Systems'' hypothesis \citep{barnes04}. The candidate planet we have identified has $M \sin i = 1.3\ M_{Sat}$ and $a$ = 1.0 AU, which is very similar to the properties of their stable test planet. However, the simulations presented by \citet{raymond05} were based on orbital parameters for the previously known planets that are somewhat different than we have determined based on the new radial velocity data and we find a slightly higher eccentricity for the new third planet than they considered. Nevertheless, the results of their robust simulations support our proposal that the third planet exists and that the orbital parameters we have derived for it are plausible. 

\section{DISCUSSION}
\begin{figure}
\epsscale{0.53}
\plotone{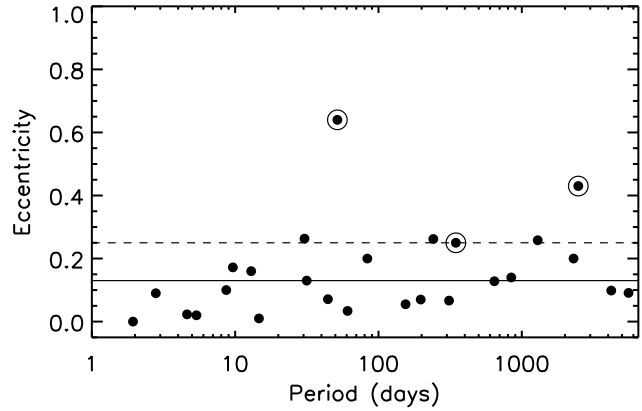}
\caption{Orbital eccentricity as a function of period for the planets in the eight systems containing three or more planets (filled circles, see Table~\ref{tab:table4} for the data and sources). The median value for this sample is 0.13 (indicated by the solid line), while the median orbital eccentricity of all the known exoplanets is 0.25 (indicated by the dashed line). The planets in the HD 74156 system (open circles) are notably eccentric relative to the planets in the systems with three or more planets. Note that the orbital eccentricity of HD 74156 d (P = 347 days) is poorly constrained with the current data.}
\label{fig:f6}
\end{figure}

With our detection of a third planet around HD 74156, it becomes the eighth normal star to host three or more planets. Currently, five other stars are known to host three planets (GJ 876, $\upsilon$ And, GJ 581, HD 69830, and HD 37124) and two are known to host four planets ($\rho^{1}$ Cnc and $\mu$ Arae). The most recently determined orbital eccentricities and periods for these planets are tabulated along with the source of the data in Table~\ref{tab:table4}. These same data are plotted in Figure~\ref{fig:f6}. The planets in systems containing three or more planets have a median orbital eccentricity of 0.13, which is significantly lower than the median of 0.25 for all known exoplanets. This could be due to the increased likelihood of dynamical instabilities to develop in higher order planetary systems where one or more component has a significantly elliptical orbit. Indeed, \citet{chat07} have proposed that dynamical instabilities leading to planet ejection can quickly arise in systems initially containing three gas giants without requiring special orbital configurations or additional evolutionary mechanisms. It might then be expected that the three planet systems which survive for the timescale of the typical known exoplanet host star's age (i.e. those that are detectable with current techniques) would be those with planets having low orbital eccentricities. 

This hypothesis seems to be supported by the sample of previously known systems containing three or more planets. However, HD 74156 b and c are in orbits that are much more eccentric than any of the components in these systems. Is HD 74156 a rare example of a system that survived a period of dynamical instability with three gas giants in highly eccentric orbits, or does its existence point to other mechanisms as playing an important role in the evolution of planetary systems? Studies of the dynamical characteristics of HD 74156 under the assumption of the previous two planet model have been carried out by \citet{nagasawa03}, \citet{barnes04}, \citet{raymond05}, \citet{raymond06}, \citet{adams06}, \citet{barnes06}, \citet{libert06a}, and \citet{libert06b}. Clearly, new dynamical studies should be undertaken now that we have uncovered the third planet in order to understand the possible origins of this system's unique configuration. 

\acknowledgments
We would like to thank Eugenio Rivera, Stefano Meschiari, Aaron Wolf, and Greg Laughlin for providing the \textit{Systemic Console} to the community. We also thank an anonymous referee for a careful reading of the manuscript and constructive comments which improved this paper. We acknowledge support from NASA GO-09407, GO-09969, GO-09971, GO-10103, GO-10610, and GO-10989 from the Space Telescope Science Institute, which is operated by the Association of Universities for Research in Astronomy, Inc., under NASA contract NAS5-26555; and from JPL 1227563 ({\it SIM} MASSIF Key Project, Todd Henry, P.I.), administered by the Jet Propulsion Laboratory.

\clearpage
\begin{deluxetable}{lr}
\tabletypesize{\scriptsize}
\tablecolumns{2}
\tablewidth{0pc}
\tablecaption{HET Radial Velocities for HD 74156}
\tablehead{
 \colhead{HJD - 2450000.0} &
 \colhead{RV (m s$^{-1}$)} 
}
\startdata
3342.8969 &  -47.28 $\pm$  3.84 \\
3347.0019 &  -41.26 $\pm$  2.58 \\
3355.8330 &  -34.26 $\pm$  2.89 \\
3357.8444 &  -35.63 $\pm$  2.91 \\ 
3359.8504 &  -28.42 $\pm$  4.95 \\
3360.9748 &  -37.04 $\pm$  3.19 \\
3364.9763 &  -41.36 $\pm$  3.23 \\
3365.8259 &  -40.73 $\pm$  2.89 \\
3367.8213 &  -64.14 $\pm$  2.93 \\
3383.9213 &  -64.01 $\pm$  3.37 \\
3390.7526 &  -27.69 $\pm$  2.82 \\
3448.7387 &   33.62 $\pm$  1.59 \\
3451.7306 &   27.48 $\pm$  2.41 \\
3476.6500 & -118.87 $\pm$  2.57 \\
3480.6372 & -127.55 $\pm$  5.68 \\
3481.6318 &  -82.68 $\pm$  2.08 \\
3482.6317 &  -57.03 $\pm$  2.13 \\
3664.9959 &  141.47 $\pm$  1.96 \\
3675.9704 &  127.44 $\pm$  2.24 \\
3676.9844 &  123.13 $\pm$  3.69 \\
3682.9515 &   20.69 $\pm$  2.78 \\
3687.9321 &   15.18 $\pm$  2.50 \\
3689.9285 &   61.28 $\pm$  2.54 \\
3691.9156 &   92.73 $\pm$  2.98 \\
3697.9126 &  123.87 $\pm$  2.03 \\
3703.8858 &  147.52 $\pm$  4.60 \\
3708.8817 &  146.26 $\pm$  3.24 \\
3710.8789 &  150.70 $\pm$  2.58 \\
3718.0141 &  136.36 $\pm$  4.98 \\
3724.8303 &  129.03 $\pm$  3.12 \\
3728.8252 &  121.82 $\pm$  2.94 \\
3731.9672 &   83.91 $\pm$  3.86 \\
3733.8020 &   39.79 $\pm$  3.20 \\
3734.8087 &   -3.84 $\pm$  2.64 \\
3736.9455 &  -86.32 $\pm$  3.54 \\
3741.7833 &   60.98 $\pm$  2.69 \\
3742.7830 &   73.14 $\pm$  2.33 \\
3743.7904 &   83.45 $\pm$  2.72 \\
3748.7726 &  128.14 $\pm$  3.04 \\
3751.7677 &  139.71 $\pm$  3.00 \\
3753.7634 &  150.89 $\pm$  2.61 \\
3754.7496 &  141.44 $\pm$  2.43 \\
3756.7485 &  141.78 $\pm$  3.11 \\
3764.7353 &  140.77 $\pm$  2.69 \\
3832.6754 &  100.92 $\pm$  2.17 \\
3833.6962 &   94.87 $\pm$  2.55 \\
3834.6755 &   85.38 $\pm$  1.92 \\
3835.6671 &   63.55 $\pm$  2.01 \\
3838.6638 &  -35.39 $\pm$  1.96 \\
3841.6443 &  -46.03 $\pm$  4.92 \\
3845.6318 &   72.93 $\pm$  2.72 \\
3846.6533 &   77.92 $\pm$  2.69 \\
4029.9895 &  134.42 $\pm$  2.58 \\
4035.9910 &  116.77 $\pm$  3.07 \\
4038.9782 &  102.21 $\pm$  1.65 \\
4039.9707 &  102.13 $\pm$  2.46 \\
4040.9590 &   79.20 $\pm$  2.02 \\
4043.9637 &   11.48 $\pm$  1.25 \\
4044.9538 &  -18.63 $\pm$  1.81 \\
4045.9556 &  -72.29 $\pm$  2.05 \\
4050.9539 &   42.33 $\pm$  2.30 \\
4051.9467 &   52.82 $\pm$  2.67 \\
4052.9393 &   67.80 $\pm$  1.72 \\
4073.8806 &  125.65 $\pm$  1.92 \\
4079.8634 &  130.08 $\pm$  2.64 \\
4087.8437 &   95.94 $\pm$  1.78 \\
4106.7835 &   63.79 $\pm$  2.55 \\
4109.7887 &   93.54 $\pm$  2.53 \\
4110.7994 &  106.19 $\pm$  2.29 \\
4129.8706 &  105.98 $\pm$  2.14 \\
4130.7409 &  103.57 $\pm$  2.36 \\
4133.8470 &   88.55 $\pm$  2.69 \\
4134.7264 &   92.38 $\pm$  2.58 \\
4135.8673 &  103.20 $\pm$  2.85 \\
4136.8409 &   90.93 $\pm$  3.26 \\
4148.6771 &  -81.12 $\pm$  3.41 \\
4156.6604 &   48.73 $\pm$  3.71 \\
4159.7804 &   66.51 $\pm$  2.32 \\
4166.7627 &   89.64 $\pm$  2.10 \\
4167.7585 &   85.27 $\pm$  2.00 \\
4211.6299 &   67.62 $\pm$  2.47 \\
4231.6001 &  104.84 $\pm$  1.78 \\
\enddata
\label{tab:table1}
\end{deluxetable}

\clearpage
\begin{deluxetable}{lccc}
\tabletypesize{\scriptsize}
\tablecolumns{4}
\tablewidth{0pc}
\tablecaption{The Radial Velocity Samples}
\tablehead{
 \colhead{Sample} &
 \colhead{Time Span} &
 \colhead{N} &
 \colhead{RMS (m\ s$^{-1}$)\tablenotemark{a}}
}
\startdata
 ELODIE  & 1998.02 -- 2003.37 & 48 &  9.5 \\
 CORALIE & 2001.03 -- 2003.74 & 38 & 11.1 \\
 HET     & 2004.92 -- 2007.36 & 82 &  6.0 \\
\enddata
\tablenotetext{a}{RMS of the residuals from the Robust three planet model.}
\label{tab:table2}
\end{deluxetable}

\clearpage
\begin{deluxetable}{lccc}
\tabletypesize{\scriptsize}
\tablecolumns{4}
\tablewidth{0pc}
\tablecaption{Derived Parameters for the HD 74156 Planets}
\tablehead{
 \colhead{Parameter} &
 \colhead{HD 74156 b} &
 \colhead{HD 74156 c} &
 \colhead{HD 74156 d}
}
\startdata
\sidehead{Robust solution}
$P$ (days)      & 51.65 $\pm$ 0.01 & 2476.7 $\pm$ 8.7 & 346.6 $\pm$ 3.6 \\
$K$ (m s$^{-1}$)& 115.1 $\pm$ 1.5 & 115.5 $\pm$ 2.3 & 10.5 $\pm$ 1.2    \\
$T_{0}$ (HJD-2450000.0) & 1980.8 $\pm$ 0.1  & 952.2 $\pm$ 13.8 & 678.2 $\pm$ 44.2 \\
$e$             & 0.64 $\pm$ 0.01 & 0.43 $\pm$ 0.01 & 0.25 $\pm$ 0.11    \\
$\omega$ ($\deg$) & 175.8 $\pm$ 0.8 & 261.3 $\pm$ 2.0 & 166.5 $\pm$ 27.4 \\
$M \sin i$ ($M_{Jup}$)\tablenotemark{a}& 1.88 $\pm$ 0.03 & 8.03 $\pm$ 0.12 & 0.40 $\pm$ 0.02 \\
\sidehead{Least Squares solution}
$P$ (days)      & 51.65 $\pm$ 0.01 & 2481.8 $\pm$ 8.5 & 339.6 $\pm$ 2.6 \\
$K$ (m s$^{-1}$)& 114.9 $\pm$ 1.8 & 118.1 $\pm$ 2.8 & 13.5 $\pm$ 2.0    \\
$T_{0}$ (HJD-2450000.0) & 1980.8 $\pm$ 0.1  & 927.6 $\pm$ 14.3 & 757.5 $\pm$ 25.7 \\
$e$             & 0.64 $\pm$ 0.01 & 0.44 $\pm$ 0.01 & 0.55 $\pm$ 0.07    \\
$\omega$ ($\deg$) & 175.8 $\pm$ 0.8 & 257.6 $\pm$ 2.3 & 188.1 $\pm$ 11.1 \\
$M \sin i$ ($M_{Jup}$)\tablenotemark{a}& 1.87 $\pm$ 0.03 & 8.19 $\pm$ 0.13 & 0.45 $\pm$ 0.04 \\
\enddata
\tablenotetext{a}{Assuming $M_{\star}$ = 1.24 $\pm$ 0.04 $M_{\sun}$ \citep{takeda07}.}
\label{tab:table3}
\end{deluxetable}

\clearpage
\begin{deluxetable}{cccc}
\tabletypesize{\scriptsize}
\tablecolumns{4}
\tablewidth{0pc}
\tablecaption{Orbital Parameters for Planets in Systems with Three or More Planets}
\tablehead{
 \colhead{Planet} &
 \colhead{$P$ (days)} &
 \colhead{$e$} &
 \colhead{Source}
}
\startdata
HD 74156 b  &  51.65  &  0.64  &  1 \\
HD 74156 c  &  2476.7 &  0.43  &  1 \\
HD 74156 d  &  346.6  &  0.25  &  1 \\
GJ 876 b    &  60.83  &  0.03  &  2 \\
GJ 876 c    &  30.46  &  0.26  &  2 \\
GJ 876 d    &  1.94   &  0.00  &  2 \\
$\rho^{1}$ Cnc b  &  14.65  & 0.01  & 3 \\
$\rho^{1}$ Cnc c  &  44.36  & 0.07  & 3 \\
$\rho^{1}$ Cnc d  &  5552   & 0.09  & 3 \\
$\rho^{1}$ Cnc e  &  2.80   & 0.09  & 3 \\
$\upsilon$ And b  &  4.62   & 0.02  & 3 \\
$\upsilon$ And c  &  241.2  & 0.26  & 3 \\
$\upsilon$ And d  &  1290.1 & 0.26  & 3 \\
GJ 581 b    &  5.37   &  0.02  &  4 \\
GJ 581 c    &  12.93  &  0.16  &  4 \\
GJ 581 d    &  83.6   &  0.20  &  4 \\
HD 69830 b  &  8.67   &  0.10  &  5 \\
HD 69830 c  &  31.56  &  0.13  &  5 \\
HD 69830 d  &  197.0  &  0.07  &  5 \\
$\mu$ Arae b&  643.25 &  0.13  &  6 \\
$\mu$ Arae c&  9.64   &  0.17  &  6 \\
$\mu$ Arae d& 310.55  &  0.07  &  6 \\
$\mu$ Arae e& 4205.8  &  0.10  &  6 \\
HD 37124 b  & 154.46  &  0.06  &  3 \\
HD 37124 c  & 2295    &  0.20  &  3 \\
HD 37124 d  & 843.6   &  0.14  &  3 \\
\enddata
\tablerefs{(1) This Paper, (2) \citet{rivera05}, (3) \citet{butler06}, (4) \citet{udry07}, (5) \citet{lovis06}, (6) \citet{pepe07}}
\label{tab:table4}
\end{deluxetable}

\clearpage

\clearpage

\clearpage

\clearpage

\clearpage

\clearpage


\newpage
\begin{thebibliography}{}

\bibitem[Adams \& Laughlin(2006)]{adams06}Adams, F., \& Laughlin, G. 2006, \apj, 649, 992

\bibitem[Barnes \& Greenberg(2006)]{barnes06}Barnes, R., \& Greenberg, R. 2006, \apj, 652, L53

\bibitem[Barnes \& Raymond(2004)]{barnes04}Barnes, R., \& Raymond, S. N. 2004, \apj, 617, 569

\bibitem[Bean et al.(2007)]{bean07}Bean, J. L., McArthur, B. E., Benedict, G. F., Harrison, T. E., Bizyaev, D., Nelan, E., \& Smith, V. V. 2007, \aj, 134, 749

\bibitem[Benedict et al.(2002)]{benedict02}Benedict, G. F., et al. 2002, \apj, 581, 115

\bibitem[Benedict et al.(2006)]{benedict06}Benedict, G. F., et al. 2006, \aj, 132, 2206

\bibitem[Bouchy et al.(2005)]{bouchy05}Bouchy, F., et al. 2005, \aap, 444, L15

\bibitem[Butler et al.(1996)]{butler96}Butler, R. P., Marcy, G. W., Williams, E., McCarthy, C., Dosanjh, P., \& Vogt, S. S. 1996, \pasp, 108, 550

\bibitem[Butler et al.(2006)]{butler06}Butler, R. P., et al. 2006, \apj, 646, 505

\bibitem[Castellano(2000)]{castellano00}Castellano, T. 2000, \pasp, 112, 821

\bibitem[Charbonneau et al.(2000)]{charbonneau00}Charboneau, D., Brown, T., M., Latham, D. W., \& Mayor, M. 2000, \apj, 529, L45 

\bibitem[Chatterjee et al.(2007)]{chat07}Chatterjee, S., Ford, E. B., \& Rasio, \apj submitted, astro-ph/0703166

\bibitem[Chiang \& Murray(2002)]{chiang02}Chiang, E., \& Murray, N. 2002, \apj, 576, 473

\bibitem[Cowan et al.(2007)]{cowan07}Cowan, N. B., Agol, E., \& Charbonneau, D. 2007, \mnras, 379, 641

\bibitem[Deming et al.(2005)]{deming05}Deming, D., Seager, S., Richardson, L. J., \& Harrington, J. 2005, Nature, 434, 740

\bibitem[Endl et al.(2002)]{endl02}Endl, M., K\"{u}rster, M., Els, S, Hatzes, A. P., Cochran, W. D., Dennerl, \& K. D\"{o}bereiner, S. 2002, \aap, 392, 671

\bibitem[ESA(1997)]{esa97}ESA 1997, The \textit{Hipparcos} and Tycho Catalogues (ESA SP-1200; Noordwijk: ESA)

\bibitem[Fischer \& Valenti (2005)]{fv05}Fischer, D. A., \& Valenti, J. 2005, \apj, 622, 1102

\bibitem[Ford et al.(2005)]{ford05}Ford, E. B., Lystad, V., \& Rasio, F. A. 2005, Nature, 434, 873

\bibitem[Gillon et al.(2007)]{gillon07}Gillon, M., et al. 2007, \aap accepted, astro-ph/0705.2219

\bibitem[Harrington et al.(2006)]{harrington06}Harrington, J., et al. 2005, Science, 314, 623

\bibitem[Henry et al.(2000)]{henry00}Henry, G. W., Marcy, G. W., Butler, R. P., \& Vogt, S. S. 2000, \apj, 529, L41

\bibitem[Janson et al.(2007)]{janson07}Janson, M., et al. 2007, \aj, 133, 244

\bibitem[Jefferys et al.(1988)]{jefferys88}Jefferys, W. H., Fitzpatrick, M. J., \& McArthur, B. 1988, Celestial Mechanics 41, 39
	
\bibitem[Knutson et al.(2007)]{knutson07}Knutson, H. A., et al. 2007, Nature, 447, 183
	
\bibitem[Laughlin \& Adams(1999)]{laughlin99}Laughlin, G., \& Adams, F. C. 1999, \apj, 526, 881

\bibitem[Libert \& Henrad(2006a)]{libert06a}Libert, A.-S., \& Henrad, J. 2006a, \aap, 461, 759

\bibitem[Libert \& Henrad(2006b)]{libert06b}Libert, A.-S., \& Henrad, J. 2006b, Icarus, 183, 186

\bibitem[L\`opez-Morales et al.(2006)]{lopez06}L\`opez-Morales, M., Morrell, N. I., Butler, R. P., \& Seager, S. 2006, \pasp, 118, 1506

\bibitem[Lovis et al.(2006)]{lovis06}Lovis, C., et al. 2006, Nature, 441, 305

\bibitem[McArthur et al.(2004)]{mcarthur04}McArthur, B. E., et al. 2004, \apj, 614, L81

\bibitem[Naef et al.(2004)]{naef04}Naef, D., Mayor, M., Beuzit, J. L., Perrier, C., Queloz, D., Sivan, J. P., \& Udry, S. 2004, \aap, 414, 351 

\bibitem[Nagasawa et al.(2003)]{nagasawa03}Nagasawa, M., Lin, D. N. C., \& Ida, S. 2003, \apj, 586, 1374

\bibitem[Pepe et al.(2007)]{pepe07}Pepe, F., et al. 2007, \aap, 462, 769 

\bibitem[Piskunov \& Valenti (2002)]{piskunov02}Piskunov, N. E., \& Valenti, J. A. 2002, \aap, 385, 1095

\bibitem[Press et al.(1992)]{press92}Press, W. H., Teukolsky, S. A., Vetterling, W. T., \& Flannery, B. P. 1992, Numerical Recipes: The Art of Scientific Computing (2nd Edition; Cambridge, U.K.: Cambridge University Press)

\bibitem[Raymond \& Barnes(2005)]{raymond05}Raymond, S. N., \& Barnes, R. 2005, \apj, 619, 549

\bibitem[Raymond et al.(2006)]{raymond06}Raymond, S. N., Barnes, R., \& Kaib, N. A. 2006, \apj, 644, 1223

\bibitem[Rey(1983)]{rey83}Rey, W. J. J. 1983, Introduction to Robust and Quasi-Robust Statistical Methods (New York: Springer-Verlag)

\bibitem[Richardson et al.(2007)]{richardson07}Richardson, L. J., Deming, D., Horning, K., Seager, S., \& Harrington, J. 2007, Nature, 445, 892

\bibitem[Rivera et al.(2005)]{rivera05}Rivera, E. J., et al. 2005, 634, 625

\bibitem[Sato et al.(2005)]{sato05}Sato, B., et al. 2005, \apj, 633, 465

\bibitem[Santos et al.(2003)]{santos03}Santos, N. C., Israelian, G., Mayor, M., Rebolo, R., \& Udry, S. 2003, \aap, 398, 363

\bibitem[Shankland et al.(2006)]{shankland06}Shankland, P. D., et al. 2006, \apj, 653, 700

\bibitem[Takeda et al.(2007)]{takeda07}Takeda, G., Ford, E. B., Sills, A., Rasio, F. A., Fischer, D. A., \& Valenti, J. A. 2007, \apjs, 168, 297

\bibitem[Tull(1998)]{tull98}Tull, R. G. 1998, Proc. SPIE, 3355, 387

\bibitem[Udry et al.(2007)]{udry07}Udry, S. et al. 2007, \aap, 469, L43

\bibitem[Valenti \& Fischer(2005)]{vf05}Valenti, J. A., \& Fischer, D. A. 2005, \apjs, 159, 141

\bibitem[Wright(2005)]{wright05} Wright, J. T. 2005, \pasp, 117, 657

\end{thebibliography}
\end{document}